# Four Distinct Pathways to the Element Abundances in Solar Energetic Particles

**Donald V. Reames**
IPST, University of Maryland, College Park, MD, USA

**Abstract** Based upon recent evidence from abundance patterns of chemical elements in solar energetic particles (SEPs), and, ironically, the belated inclusion of H and He, we can distinguish four basic SEP populations: (1) SEP1 – pure "impulsive" SEPs are produced by magnetic reconnection in solar jets showing steep power-law enhancements of $1 \leq Z \leq 56$ ions versus charge-to-mass ratio $A/Q$ from a ≈3 MK plasma. (2) SEP2 – ambient ions, mostly protons, plus SEP1 ions reaccelerated by the shock wave driven by the narrow coronal mass ejection (CME) from the same jet. (3) SEP3 – a "gradual" SEP event is produced when a moderately fast, wide CME-driven shock wave barely accelerates ambient protons while preferentially accelerating accumulated remnant SEP1 ions from an active region fed by multiple jets. (4) SEP4 – a gradual SEP event is produced when a very fast, wide CME-driven shock wave is completely dominated by ambient coronal seed population of 0.8 – 1.8 MK plasma usually producing a full power law vs. $A/Q$ for $1 \leq Z \leq 56$ ions. We begin with element abundances in the photosphere that are fractionated during transport up to the corona based upon their first ionization potential (FIP); this important "FIP effect" for SEPs provides our reference abundances and is different for SEPs from that for the solar wind. We then show evidence for each of the processes of acceleration, reacceleration, and transport that conspire to produce the four abundances patterns we distinguish.

## 1 Introduction

Relative abundances of the chemical elements have been a key to understanding the physics in many astrophysical settings. Solar energetic particles (SEPs) are no exception, and for these SEPs, abundance measurements continue to produce surprising new results. The first evidence for the existence of SEPs was reported by Forbush (1946) who observed what we now call ground level events (GLEs) produced when SEPs at GeV energy fragment in nuclear cascades through the atmosphere to produce secondary particles seen at ground level. Of course, these GLEs provide no information at all on the incoming SEP abundances. Direct measurements of SEP abundances were made when nuclear emulsions flown on sounding rockets from Fort Churchill, Manitoba first measured C, N, and O (Fichtel and Guss 1961) and eventually Fe (Bertsch et al. 1969). Measurements have progressed for individual elements up through Fe (Teegarden et al. 1973; McGuire et al. 1979; Cook et al. 1984; Meyer 1985; Reames 1995a, 2014, 2017a, 2018a; Mewaldt et al. 2002) and groups of elements throughout the rest of the periodic table (Reames 2000; Mason et al. 2004; Reames and Ng 2004; Reames et al. 2014a). Yet, surprising new patterns have begun to emerge only recently as we extended abundances *down* to H (Reames 2019b, 2019c, 2019d).

Suggestions for the physical source of SEPs evolved slowly. Originally SEPs were believed to be accelerated somehow in solar flares, the only associated feature then visible on the Sun. But how could point-source flares distribute SEPs across magnetic fields over more than half of the heliosphere? Imaginative schemes such as the "bird cage" model of Newkirk and Wenzel (1978), allowing ions to hop from loop to loop across the face of the Sun, were taken quite seriously. After many years, Kahler et al. (1984) showed that SEPs had a strong (96%) association with wide, fast coronal mass ejections (CMEs), only discovered a few years before, and thus with the extensive shock waves that they drive out from the Sun. The review by Gosling (1993) entitled "The Solar Flare Myth" highlighted the emerging modern understanding that SEPs in the large, persistent "gradual" SEP events were accelerated along the broad fronts of expanding, CME-driven shock waves (e.g. Reames 1995b, 1999, 2013, 2015, 2017a, 2019e). Shock acceleration is relatively well understood (Bell 1978; Lee 1983, 2005; Jones and Ellison 1991; Zank et al. 2000, 2007; Sandroos and Vainio 2007; Ng and Reames 2008) and this model has advanced considerably (Kahler, 1992, 1994, 2001; Reames et al. 1996, 1997; Cliver, et al. 2004; Cliver and Ling, 2007; Gopalswamy et al. 2012; Mewaldt et al. 2012; Lee, et al. 2012; Cliver 2016; Desai and Giacalone 2016), including correlated spatial studies of CMEs and SEPs (Rouillard et al. 2011, 2012, 2016; Koulomvakos et al. 2019) and SEP onset timing (Tylka et al. 2003; Reames 2009a, 2009b).

From the earliest element abundance measurements of SEPs in these large, gradual events, there were attempts to compare them with abundances in the solar corona and photosphere, where measurements were also evolving slowly with time. The relative enhancement of SEP and coronal abundances was found to depend upon the first ionization potential (FIP) of the elements (Webber 1975; Meyer 1985). Elements with low FIP (below ~10 eV), which are ionized in the photosphere and chromosphere, are enhanced by a factor of nearly 3 over high-FIP elements, which begin their outward journey as neutral atoms. All elements become highly ionized in the 1-MK corona. The solar "FIP effect" is now believed to result because the ions are preferentially conveyed upward by the ponderomotive force of Alfvén waves (Laming 2015) which affect ions but not neutral atoms. However, we have found that the underlying FIP pattern of SEPs differs from that of the solar wind (Mewaldt et al. 2002; Reames 2018a).

Meyer (1985) found that all SEP events showed evidence of the FIP effect while individual events showed his "mass effect" or additional enhancement or suppression of abundances that depended upon the mass-to-charge ratio $A/Q$ of the ions. Breneman and Stone (1985) showed this to be a power law in $A/Q$ for some events, with $Q$ based upon the ionization-state measurements of Luhn et al. (1984). We can now understand this as a power-law dependence upon the magnetic-rigidity of ion scattering during transport out from the Sun (Parker 1963; Ng et al. 2003). If we compare abundances of different ions at the same velocity (or energy per nucleon), as we always do, this rigidity dependence becomes simply a relative dependence upon $A/Q$. For example, if Fe scatters less than O, then Fe/O will become enhanced early in an event and depleted later; solar rotation can then turn this time dependence into longitude dependence. Of course, $Q$ values are a function of the effective source plasma temperature $T$, so the power-law dependence of enhancements on $A/Q$ can be used to determine a best-fit value of $T$ for each event (Reames 2016a, 2016b) using theoretical values of $Q$ vs. $T$ (e.g. Mazzotta et al. 1998). While straightforward, the technique of using element abundances vs. $A/Q$ to derive source plasma temperatures is also explained in reviews by Reames (2017a, 2018b).

A surprising class of SEP events, the small "impulsive" events, emerged when measurements began to find large enhancements in $^3$He/$^4$He, relative to $(4.08 \pm 0.25) \times 10^{-4}$ in the solar wind (e.g. Gloeckler and Geiss 1998). At first, these enhancements were considered as possible evidence of nuclear fragmentation, as had been seen in galactic cosmic rays, but then Serlemitsos and Balasubrahmanyan (1975), for example, found a huge value of $^3$He/$^4$He = 1.52 $\pm$ 0.10, with no evidence of $^2$H or secondary elements Li, Be, or B. Limits on Be/O or B/O in SEP events are $< 2 \times 10^{-4}$ (e.g. McGuire et al. 1979; Cook et al. 1984). Thus fragmentation was completely excluded and a resonance process was required. Early proposals of wave-particle interactions could produce selective heating of $^3$He but failed to suggest a mechanism of acceleration. Then, Temerin and Roth (1992) suggested that copious non-relativistic electrons, observed to be streaming from impulsive SEP events (Reames et al. 1985) and producing radio type III bursts (Reames and Stone 1986), could also generate electromagnetic ion-cyclotron waves, near the gyrofrequency of $^3$He, that drove the acceleration. Separately, Liu et al. (2006) were able to use stochastic acceleration to fit the complex spectra of $^3$He and $^4$He.

Almost as spectacular as the increases in $^3$He were the enhancements in heavy elements which were found to increase all the way from He up to 1000-fold enhancements of heavy elements near Au or Pb (Reames 2000; Mason et al. 2004; Reames and Ng 2004; Reames et al. 2014a). These enhancements of the heavy elements, rising as a power law in $A/Q$, appear to be accelerated in magnetic reconnection regions, according to particle-in-cell simulations (e.g. Drake et al. 2009). The study of these increases as an indication of source-plasma temperatures grew slowly. Modest increases were first observed in Fe/O and elements up to Fe, in $^3$He-rich events. Reames et al.

(1994) noticed that He, C, N, and O were relatively unenhanced in impulsive events, Ne, Mg, and Si, were enhanced by a factor of 2.5 and Fe was enhanced by a factor of ≈7. They noted that at a temperature of 3–5 MK, He, C, N, and O would be fully ionized, or nearly so, Ne, Mg, and Si would all be in a similar stable state with two orbital electrons so that *Q/A* ≈ 0.43, and Fe would have *Q/A* ≈ 0.28, giving a nascent power law in *Q/A*, or *A/Q*. When heavier elements were observed, this power law was subsequently extended across the periodic table at *T* ≈ 3 MK as a 3.64 ± 0.15 power of *A/Q*, on average (Reames et al. 2014a). When we fit abundance enhancements of individual impulsive SEP event as power-laws vs. *A/Q*, we try *A/Q* values at many values of *T* and select the value of *T* giving the smallest $\chi^2$. Nearly all impulsive events fall in the 2.5–3.2 MK region (Reames et al. 2014b, 2015). Improving accuracy shows that enhancements in impulsive SEP events increase with Ne > Mg > Si, opposite the *Z* order; this order of *A/Q* is unique for the *T* ≈ 3 MK region.

While gradual SEP events were associated with wide, fast CMEs, impulsive SEP events either had no significant CME or narrow CMEs (Kahler et al. 2001) which led them to be associated with solar jets, and the events have subsequently been associated directly with solar jets (Bučík et al. 2018a, 2018b). Early measurements had shown that the charge state of Fe in impulsive events was 20.5±1.2 (Luhn et al. 1987) while that for gradual events varied from 11 to 15 (Luhn et al 1987, Mason et al 1995, Leske et al. 1995, 2001; Tylka et al 1995; Klecker 2013), over a wide range of energies. However, subsequent measurements in impulsive events showed that *Q*-values varied with energy, probably because of stripping to an equilibrium charge state by passage through a small amount of material after acceleration at a depth found to be ≈1.5 solar radii (DiFabio et al. 2008). This increases the importance of estimates of *T* = 2.5–3.2 MK obtained from abundances in impulsive events (Reames et al. 2014b, 2015) since these temperatures and the corresponding ionization states were determined at the time of acceleration, before stripping. The strong power-law abundances imparted during acceleration are sufficiently persistent to dominate any weaker effects of later transport with altered ionization states.

For a time it seemed that impulsive and gradual SEP events could be distinguished by the presence or absence of $^3$He alone, since $^3$He seemed to be a signature of an impulsive event, but new complexity emerged when Mason et al. (1999) found a small but significant increase in $^3$He in large SEP events that would otherwise qualify as being gradual. Shock waves could easily reaccelerate residual suprathermal ions from earlier impulsive SEP events, perhaps even preferentially, since higher-velocity ions could more easily overtake the shock from downstream, especially in cases where the magnetic field lies near the plane of the shock (Desai et al 2003; Tylka et al 2001, 2005; Tylka and Lee 2006). For as many as 25% of gradual events, reacceleration of the 3-MK impulsive suprathermal ions actually *dominates* the abundances of the elements with *Z* > 2. These events seem to involve weaker shocks and perhaps quasi-perpendicular shocks, where the angle between the field ***B*** and the shock normal $\theta_{Bn} > 60°$, so that thermal ions downstream would have difficulty in overtaking the shock in order to scatter back and forth to gain energy (Tylka et al. 2005; Tylka and Lee 2006; Reames 2019d).

It is only when we include the abundance of the element H, which significantly extends the span of *A/Q*, that we find revealing upward breaks in the power-law dependence that could be direct evidence of two-component seed population for shock acceleration, at least for weak shocks, and is therefore a signature of shock-acceleration itself (Reames 2019b, 2019c, 2019d). The intermediate energies we study allow us to exploit the simple power-law behavior in *A/Q*, avoiding high-energy spectral breaks (e.g. Mewaldt et al. 2012), for example, to inter-compare a wide range of sizes and strengths of gradual and impulsive SEP events.

In this review we suggest a broad overall organizational pattern affecting the origin of element abundances in SEP events from the photosphere outward, indicating the dominant processes that modify abundances, and showing examples to justify each type of suggested behavior. Perhaps this new organization can provide a more complete framework for our understanding and for guiding future studies.

Measurements of element abundances in this article are from the *Low-Energy Matrix Telescope* (LEMT) onboard the *Wind* spacecraft, near Earth (von Rosenvinge *et al.* 1995; see also Chapt. 7 of Reames, 2017a). LEMT measures elements from H through Pb in the 2–20MeV amu$^{-1}$ region, although energies of H are limited to 2 – 2.5 MeV and LEMT resolves only element groups above Fe as shown by Reames (2000, 2017a). As a basis, we consider impulsive events listed and studied by Reames, Cliver, and Kahler (2014a, 2014b) and the gradual events listed and studied by Reames (2016a). CME data used are from the *Large Angle and Spectrometric Coronagraph* (LASCO) onboard the *Solar and Heliospheric Observatory* (SOHO) reported in the SOHO/LASCO CME catalog (Gopalswamy *et al.* 2009; https://cdaw.gsfc.nasa.gov/CME_list/).

## 2 Physical Processes that Control SEP Abundances

Figure 1 provides a roadmap for our discussion of the way physical processes combine to determine the element abundances in SEP events. From the abundances of the solar photosphere, we begin with the FIP effect after location (a) in Fig. 1, which is different for SEPs and the solar wind. Dominant acceleration mechanisms occur in magnetic reconnection regions or at CME-driven shock waves after (b). After Fig. 1(c) open field lines in jets allow SEPs to escape to produce SEP1 ions, or later, the shock-reaccelerated SEP2 ions after (e), while the corresponding ions in flares are trapped. After Fig. 1(d) shock properties determine the mix of SEP1, 2, ions and ambient corona in gradual SEP events: SEP3, where SEP1 seed ions dominate, and SEP4, where they do not. Transport can impose an *A/Q* modification at (f), more strongly in intense gradual events with self-generated waves than in scatter-free impulsive events. Examples of these SEP ions will be exhibited and discussed in the remainder of this article.

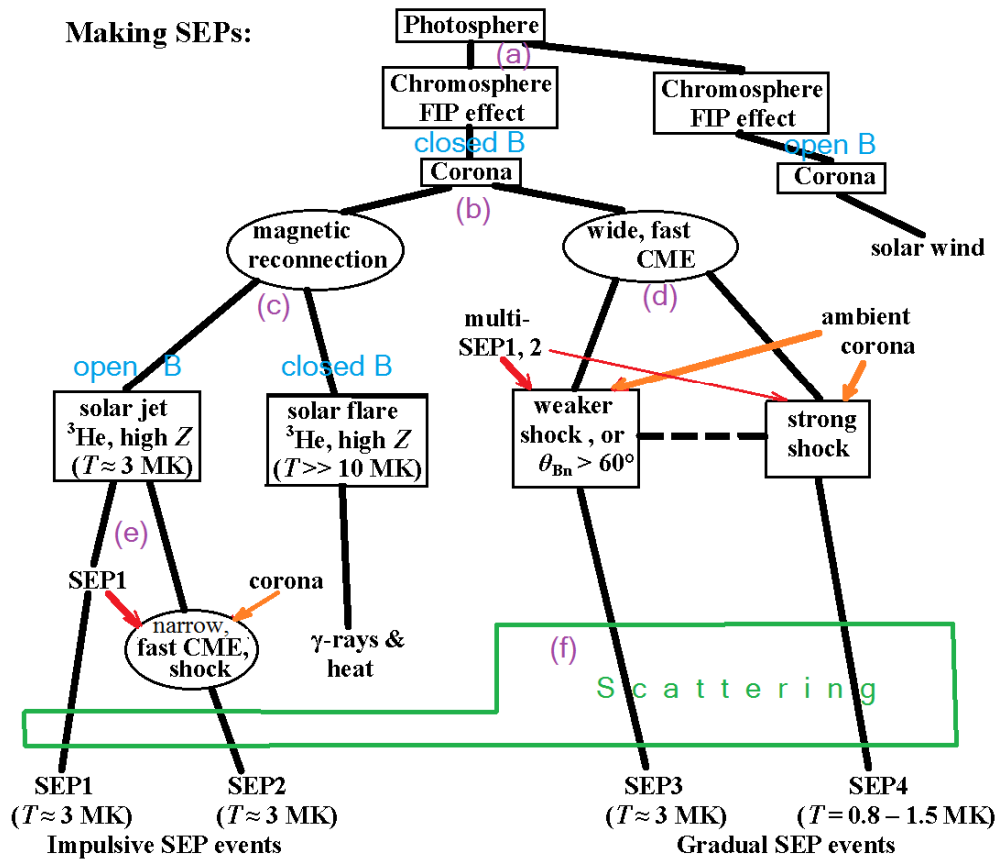

**Fig. 1** The schematic shows the locations and physical processes leading to the acceleration of SEP events with four different patterns discussed in this article. **(a)** Elements experience different FIP processing on open and closed fields, B. **(b)** SEP acceleration can occur in magnetic reconnection regions or by shock acceleration. **(c)** SEPs with $^3$He- and high-Z-rich ions escape from open B in jets, but not flares. **(d)** Weaker shocks prefer pre-accelerated SEP1 suprathermal seed ions while strong shocks deeply sample ambient corona. **(e)** Fast CMEs in jets reaccelerate the SEP1 ions to produce SEP2 **(f)** Rigidity-dependent scattering modifies abundances, especially in gradual events where high intensities produce self-generated waves.

## *2.1 Entering the Corona: The FIP Effect*

The making of the element abundance patterns of SEPs begins early when those elements are transported across the chromosphere into the corona. Figure 2(a) shows average gradual SEP abundances divided by photospheric abundances (Caffau et al. 2011; Lodders et al. 2009; see also Asplund et al. 2009) compared with recent calculations of Laming et al. (2019) for loop-like structures. See abundances listed in Table 1.

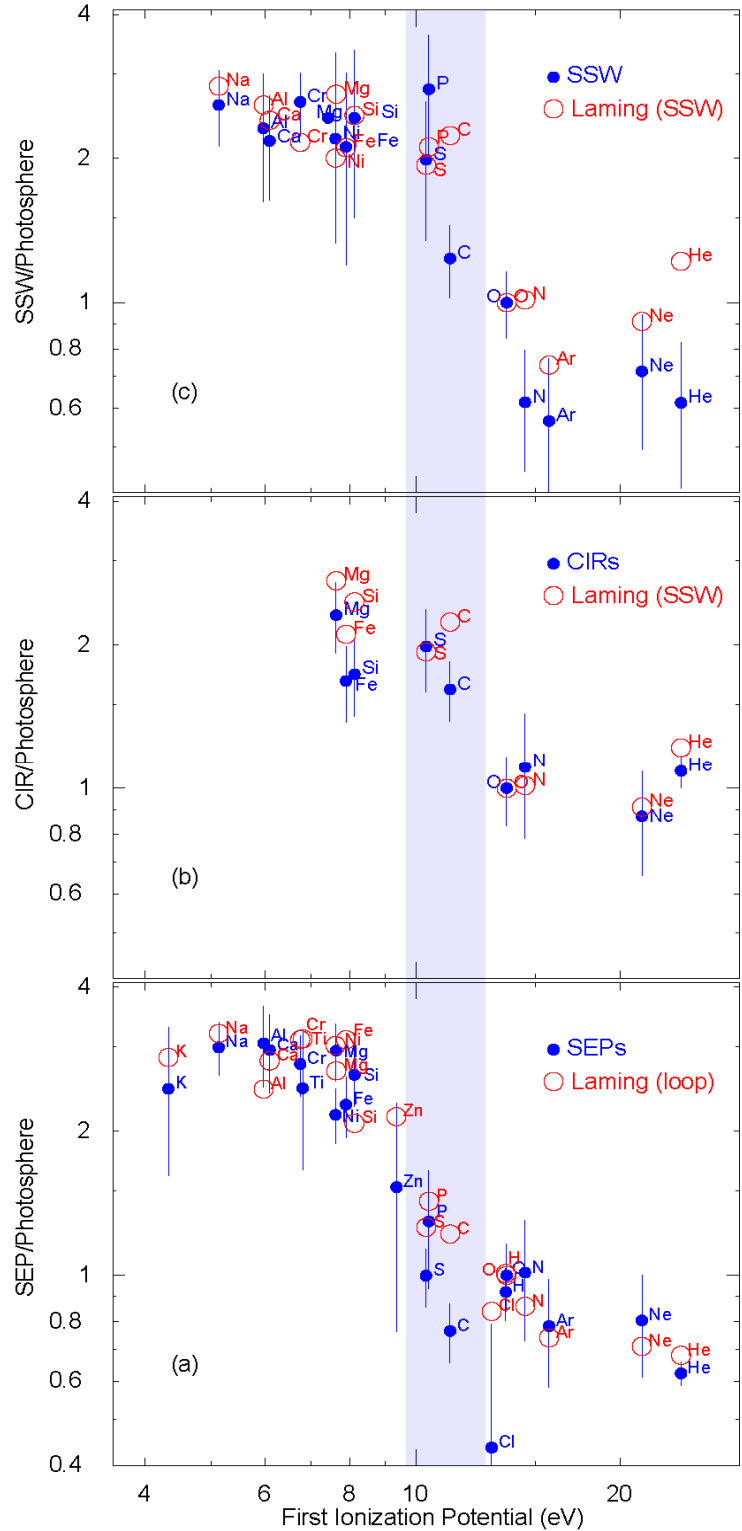

**Fig. 2** Average abundances of (**a**) SEPs, (**b**) CIR ions and (**c**) slow solar wind (SSW; Bochsler 2009), relative to solar photospheric abundances, (*solid blue*) are shown as a function of the FIP of each element and compared with theoretical calculations (*open red*) by Laming et al. (2019) for loop structures (**a**), and for the SSW (**b** and **c**). All abundances are normalized at O. The *light blue* band compares elements C, P, and S that are suppressed like high-FIP elements in SEPs, but are elevated like low-FIP elements in the solar wind. However, decreasing the photospheric C/O ratio by 20% would greatly improve the agreement of observation and theory for all three samples: SEPs, CIRs, and SSW.

It has been well established that SEP abundances differ from those of the solar wind (Mewaldt et al. 2002; Desai et al. 2003; Kahler and Reames 2003; Kahler et al. 2009; Reames 2018a, 2018b). The elements C, P, and S behave as high-FIP elements, suppressed in SEPs as shown in Fig. 2(a), but are elevated like low-FIP elements in the solar wind in Fig. 2(b) and 2(c) (Reames 2018a, 2018b; Laming et al. 2019). In terms of the theory in which the ponderomotive force of Alfvén waves enhances low-FIP elements (Laming 2015; Laming et al. 2019), Alfvén waves resonate with the loop length in closed loops resulting in reduced C, P, and S abundances in SEPs, but cannot do so for open fields leading to solar wind (Reames 2018a, 2018b; Laming et al. 2019). While we have not shown the fast (coronal hole) solar wind, the abundances of C and S are identical for the fast and slow (interstream) wind (Bochsler 2009; P is not listed).

A sample of solar wind abundances is also provided by particles accelerated at shock waves formed in corotating interaction regions (CIRs) where high-speed solar-wind streams overtake slow wind emitted earlier in the solar rotation (Reames et al. 1991; Richardson 2004; Reames 2018c) and shown in Fig. 2(b). C and S are elevated like Si and Fe in the CIR ions but behave like O, N, and Ar in the SEPs. Note also that the high-FIP elements, and especially He, agree well with the theory for the CIR ions; these elements have large errors in the SSW. He/O is suppressed for SEPs but elevated for CIR ions, both in the observations and in the theory.

The distinction between the FIP patterns in SEPs and in the solar wind is important for the origin of both. According to theory, the FIP pattern depends upon the open or closed field at the base of the corona, long before that plasma will be accelerated outward as SEPs or solar wind. It is not so surprising that plasma that will later become SEPs originates in active regions, but it is surprising that these originally-closed loops contribute little to the slow solar wind. The slow wind may come from small coronal holes from which field lines are highly divergent (Wang and Sheeley 1990).

The most significant theoretical difference in Fig. 2(a) is for C (see also Table 1). C/O is $0.420 \pm 0.010$ on average in SEPs, and is $< 0.5$ in all individual SEP events, but is $0.550 \pm 0.076$ in values quoted for the photosphere (see also Asplund et al. 2009). We see no way to reconcile the difference unless the photospheric C/O ratio is actually 20% lower. A lower value of the photospheric C/O ratio would improve the agreement of observation and theory for all three samples shown in Fig. 2: SEPs, CIRs, and SSW.

It has become common to use spectral line measurements of Si and S to measure the spatial distribution of the FIP effect in the solar corona (Brooks et al. 2016; Doschek and Warren 2019). These full-Sun maps are appropriate and extremely helpful for SEPs. It turns out that the region of high Si/S seems confined to closed magnetic loops and active regions which comparisons of SEP abundances with FIP theory (Fig. 2(a); Reames 2018a; Laming et al. 2019) have told us to be the origin of SEPs. This helps to confirm the origin of plasma that will become SEPs.

Thus, the Si/S ratio is an excellent measure of FIP for SEPs (Fig 2(a)). However, the Si/S ratio is a terrible measure of FIP for the solar wind (Fig. 2(b) or 2(c)). As seen from both CIRs and the slow solar wind in Fig. 2 and Table 1, and from publications (Reames 2018a; Laming et al. 2019), S is elevated as a low-FIP element like Si in the solar wind so Si/S cannot measure the FIP bias of atoms destined to form the solar wind. What we lack are images of measures like Mg/Ne which we would expect to be elevated for the origin of both SEPs and the solar wind. One of the advantages of SEPs for coronal studies is that a large number of elements are measured at a time, spanning a large part of the FIP pattern. SEPs are an actual sample of coronal plasma.

**Table 1** Photospheric, Reference SEPs, CIR, and Slow Solar Wind Abundances.

| | Z | FIP [eV] | Photosphere[1] | SEPs[2] | CIRs[3] | Interstream Solar Wind[4] |
|---|---|---|---|---|---|---|
| H | 1 | 13.6 | $(1.74\pm0.04)\times10^{6}$ * | $(\approx1.6\pm0.2)\times10^{6}$ | $(1.81\pm0.24)\times10^{6}$ | – |
| He | 2 | 24.6 | $1.46\pm0.07\times10^{5}$ | $91000\pm5000$ | $159000\pm10000$ | $90000\pm30000$ |
| C | 6 | 11.3 | $550\pm76$* | $420\pm10$ | $890\pm36$ | $680\pm70$ |
| N | 7 | 14.5 | $126\pm35$* | $128\pm8$ | $140\pm14$ | $78\pm5$ |
| O | 8 | 13.6 | $1000\pm161$* | $1000\pm10$ | $1000\pm37$ | 1000 |
| Ne | 10 | 21.6 | $195\pm45$ | $157\pm10$ | $170\pm16$ | $140\pm30$ |
| Na | 11 | 5.1 | $3.47\pm0.24$ | $10.4\pm1.1$ | – | $9.0\pm1.5$ |
| Mg | 12 | 7.6 | $60.3\pm8.3$ | $178\pm4$ | $140\pm14$ | $147\pm50$ |
| Al | 13 | 6.0 | $5.13\pm0.83$ | $15.7\pm1.6$ | – | $11.9\pm3$ |
| Si | 14 | 8.2 | $57.5\pm8.0$ | $151\pm4$ | $100\pm12$ | $140\pm50$ |
| P | 15 | 10.5 | $0.501\pm0.046$* | $0.65\pm0.17$ | – | $1.4\pm0.4$ |
| S | 16 | 10.4 | $25.1\pm2.9$* | $25\pm2$ | $50\pm8$ | $50\pm15$ |
| Cl | 17 | 13.0 | $0.55\pm0.38$ | $0.24\pm0.1$ | – | – |
| Ar | 18 | 15.8 | $5.5\pm1.3$ | $4.3\pm0.4$ | – | $3.1\pm0.8$ |
| K | 19 | 4.3 | $0.224\pm0.046$* | $0.55\pm0.15$ | – | – |
| Ca | 20 | 6.1 | $3.72\pm0.60$ | $11\pm1$ | – | $8.1\pm1.5$ |
| Ti | 22 | 6.8 | $0.138\pm0.019$ | $0.34\pm0.1$ | – | – |
| Cr | 24 | 6.8 | $0.759\pm0.017$ | $2.1\pm0.3$ | – | $2.0\pm0.3$ |
| Fe | 26 | 7.9 | $57.6\pm8.0$* | $131\pm6$ | $97\pm11$ | $122\pm50$ |
| Ni | 28 | 7.6 | $2.95\pm0.27$ | $6.4\pm0.6$ | – | $6.5\pm2.5$ |
| Zn | 30 | 9.4 | $0.072\pm0.025$ | $0.11\pm0.04$ | – | – |
| Se–Zr | 34–40 | – | $\approx0.0118$ | $0.04\pm0.01$ | – | – |
| Sn–Ba | 50–56 | – | $\approx0.00121$ | $0.0066\pm0.001$ | – | – |
| Os–Pb | 76–82 | – | $\approx0.00045$ | $0.0007\pm0.0003$ | – | – |

[1]Lodders et al. (2009). Ratios to O of elements from Lodders et al. (2009) are taken before correction of O by Caffau et al. (2011).
* Caffau et al. (2011).
[2] Reames (1995a, 2014, 2017a).
[3] Reames, et al. (1991); Reames (1995a).
[4] Bochsler (2009).

## 3 Impulsive SEP Events

$^3$He/$^4$He would seem to be the obvious way to choose impulsive events, but what value of $^3$He/$^4$He should we require, and at what energy? This ratio varies by orders of magnitude with energy within individual events (Mason 2007) and we know that some $^3$He is reaccelerated in large gradual events (Mason et al. 1999). Probably all SEP events have some $^3$He and there is no obvious boundary. Element abundances such as Fe/O are better behaved. Bimodal abundances of Fe/O have been observed for a long time (Reames 1988, Reames and Ng 2004) and, with the improved statistics of more-sensitive instruments, they provide an obvious distinction. Figure 3(a) shows a histogram of normalized abundances of Ne/O vs. Fe/O for 19 years of 8-hr averages at the lowest energies observed by the LEMT telescope on the *Wind* spacecraft. The two peaks in this figure define periods of impulsive and gradual SEP events (Reames et al. 2014a; Reames 2015, 2018b). The resulting average abundance pattern for the 111 impulsive SEP events that were identified and listed is shown in Fig. 3(b) (Reames et al. 2014a). All of the impulsive events which we will classify as SEP1 or SEP2 were distinguished in this way.

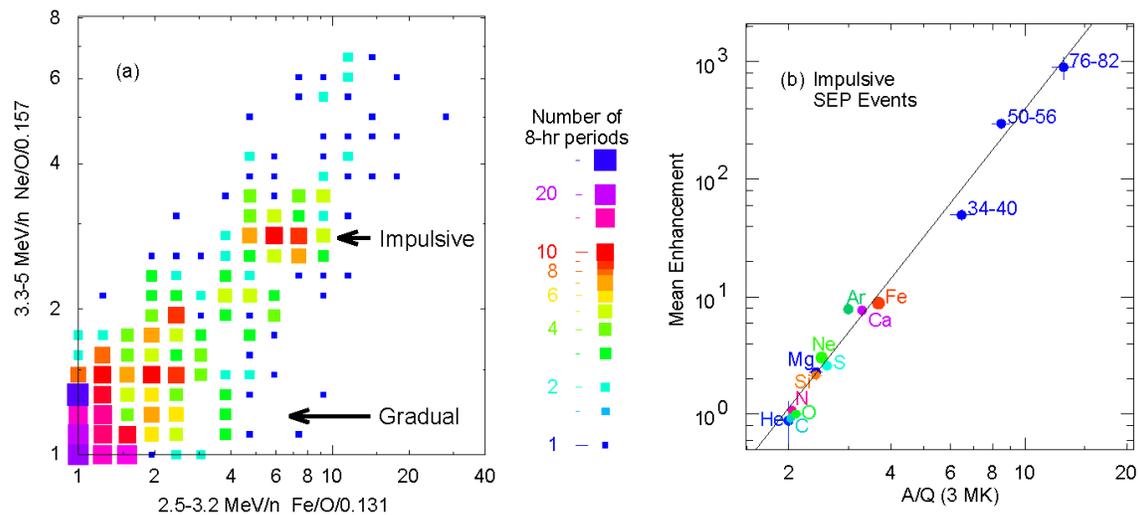

**Fig. 3. (a)** Normalized values of enhancements of Ne/O *vs.* Fe/O near 3 MeV amu$^{-1}$ are binned for all 8-hr intervals during 19 years which have errors of 20% or less. The periods near the origin at (1, 1) represents average gradual SEP event periods, that determine the normalization factors. The peak near (7, 3) defines the selection of impulsive SEP events. Candidate "impulsive" periods were chosen to have ≥4 on the abscissa. **(b)** Average enhancements of elements, relative to coronal values, are shown for the impulsive SEP events (SEP1 and SEP2) vs. *A/Q* at 3 MK (Reames et al. 2014a, Reames 2018b).

     Impulsive and gradual SEP events were once distinguished by the duration of associated X-ray bursts, but that no longer seems appropriate or relevant. It is true that most impulsive events themselves last for hours while gradual events last for days, but these durations are not completely distinct. In fact, the clearest separation of these two event classes now is that based upon bimodal distribution of abundances shown in Fig. 3(a) (Reames 1988, 2015, 2018b; Reames and Ng 2004; Reames et al. 2014a). The resulting impulsive events, thus defined, turn out to have the abundance pattern in Fig. 3(b) and the associations with narrow CMEs (Kahler et al. 2001), if any, and with jets (Bučík et al. 2018a, 2018b).

## 3.1 Power-Law Fits in A/Q and Temperature

It is important to recognize how strongly the power-law fits to the enhancements vs. *A/Q* cluster near ≈3 MK in impulsive SEP events. The region near $T \approx 3$ MK is unique because $(A/Q)_{Ne} > (A/Q)_{Mg} > (A/Q)_{Si}$ here, and the abundance enhancements turn out to be in the same order, with Ne > Mg > Si; usually the *A/Q* values tend to increase with *Z*. The characteristic enhancement of Ne/O in impulsive events (Fig. 3(a)) is a direct result of this behavior near $T \approx 3$ MK. The examples in Fig. 4 show an analysis of a cluster of impulsive SEP events. The temperatures at minimum $\chi^2/m$, and the corresponding fits of enhancement vs. *A/Q*, are shown for each event. Note how the minima of $\chi^2/m$ fall near 3 MK for these events in Fig. 4(d), as we see for nearly all impulsive SEP events (Reames et al. 2014b, 2015). Protons could not be distinguished from background for the events in Fig. 4(a) and hence were not be included in Fig. 4(e).

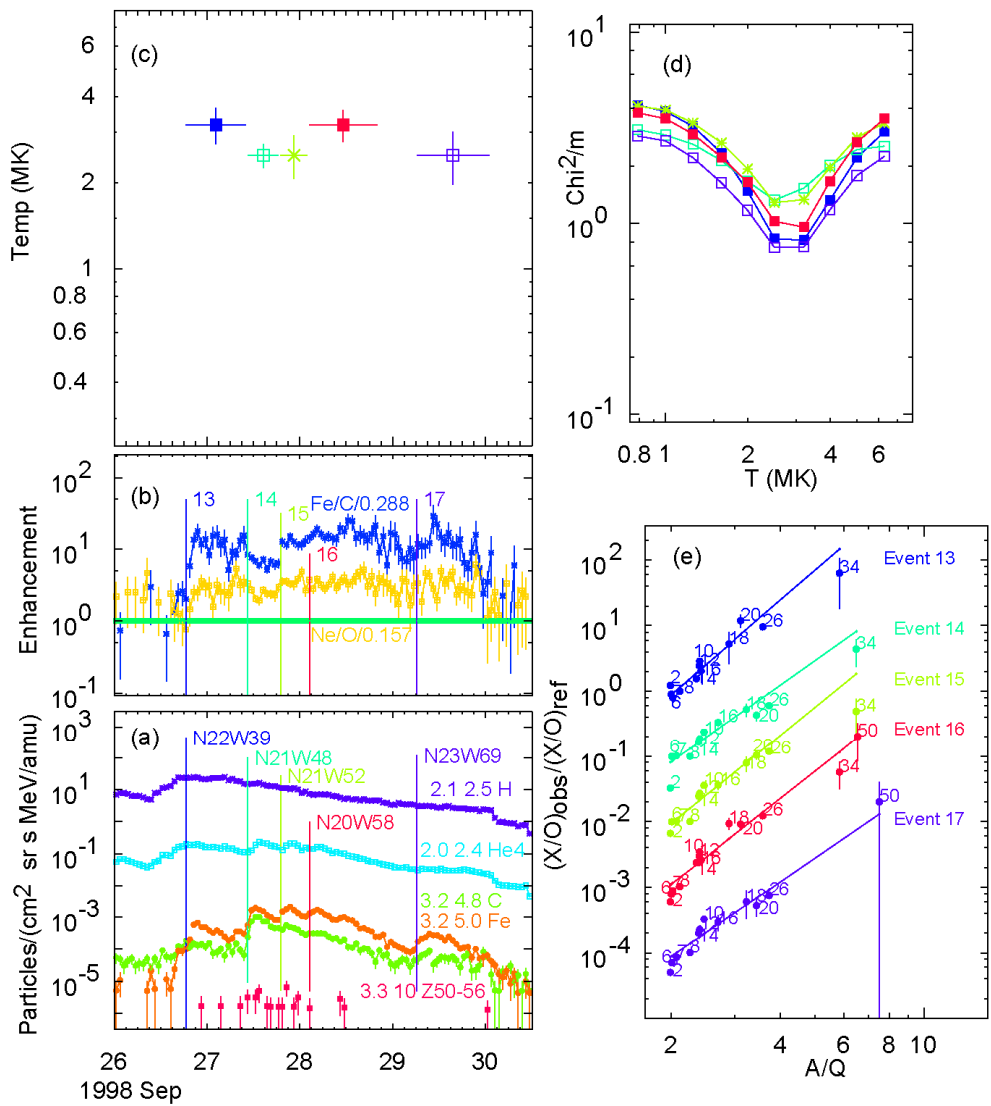

**Fig. 4** Panel **(a)** shows intensities of various ion species and **(b)** shows abundance ratios. Onset times showing event numbers in **(b)** and locations in **(a)** are color coded. Least-squares fits show $\chi^2/m$ vs. *T* (*m* is the number of degrees of freedom) in **(d)** with the best-fit *T* for each event in **(c)** and the corresponding enhancement vs. *A/Q* in **(e)**, all color coded.

## *3.2 Including H in the Abundances*

Representatives of impulsive SEP events with three different properties are shown in Fig. 5. On the left are two small simple impulsive events which we will identify as SEP1 events. Here the proton intensities are well predicted by extrapolating the fit line obtained for elements with $Z >2$. In the center is a larger event with excess protons compared with the extrapolation of the $Z > 2$ fit line that will become a SEP2 event (Reames 2019b). On the right is an impulsive SEP event with a large suppression of He compared with the fit line, but with no proton excess, which will make this event also a SEP1 event (Reames 2019a).

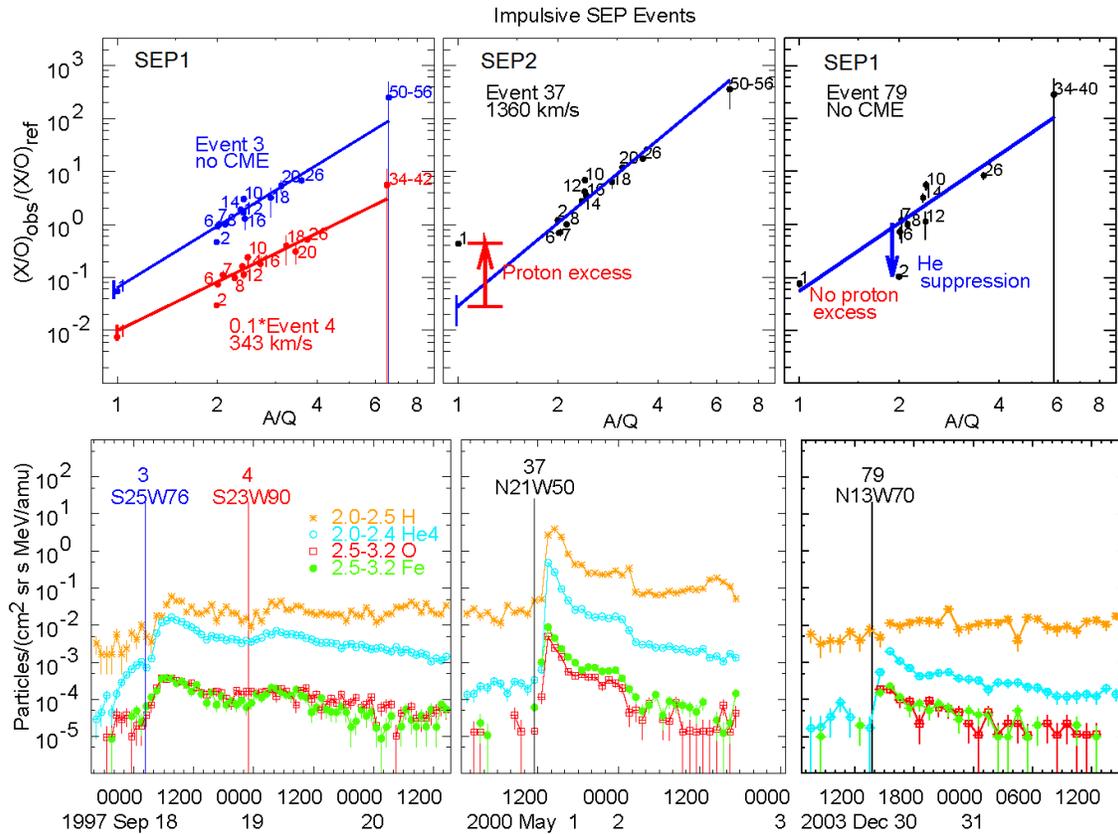

**Fig. 5** Intensities are shown *below* and enhancements vs. *A/Q above* for three periods with impulsive SEP events. Two events have "normal" proton abundances on the *left*. An event, with an associated 1360 km s$^{-1}$ CME, has a "proton excess" in the *center*. On the *right* is shown an event with "He suppression", when compared with the fit line for $Z > 2$ ions, but no proton excess.

A more-complete understanding of the proton behavior in impulsive SEP events can be found in Fig. 6. Figure 6(a) shows that the events with a significant proton excess tend to be larger events, usually with fast CMEs where CME-driven shock waves have probably re-accelerated the ions to produce SEP2. Figure 6(b) suggests possible two-component SEP2 events and Fig 6(c) shows energy spectra in a seed population where different H/O ratios of different components allow weak shock waves to select H from the corona and ions with $Z > 6$, shown as O, from pre-enhanced SEP1 ions.

**Fig. 6** Panel **(a)** shows the peak intensity of ≈2 MeV protons vs. the proton excess in impulsive SEP events with color and size of each point showing the speed of the associated CME, if any. Events with proton excesses tend to be large events with fast CMEs. Panel **(b)** shows a possible mix of shock-accelerated ions from a seed population with both impulsive (SEP1) ions and ambient coronal ions. SEP1 ions are known by their steep positive slope and $T ≈ 3$ MK. Panel **(c)** sketches the way the threshold of a weak shock can select protons from one component of the seed population and O etc. from the other. SEP1 ions have a lower H/O ratio (Reames 2019b, 2019d).

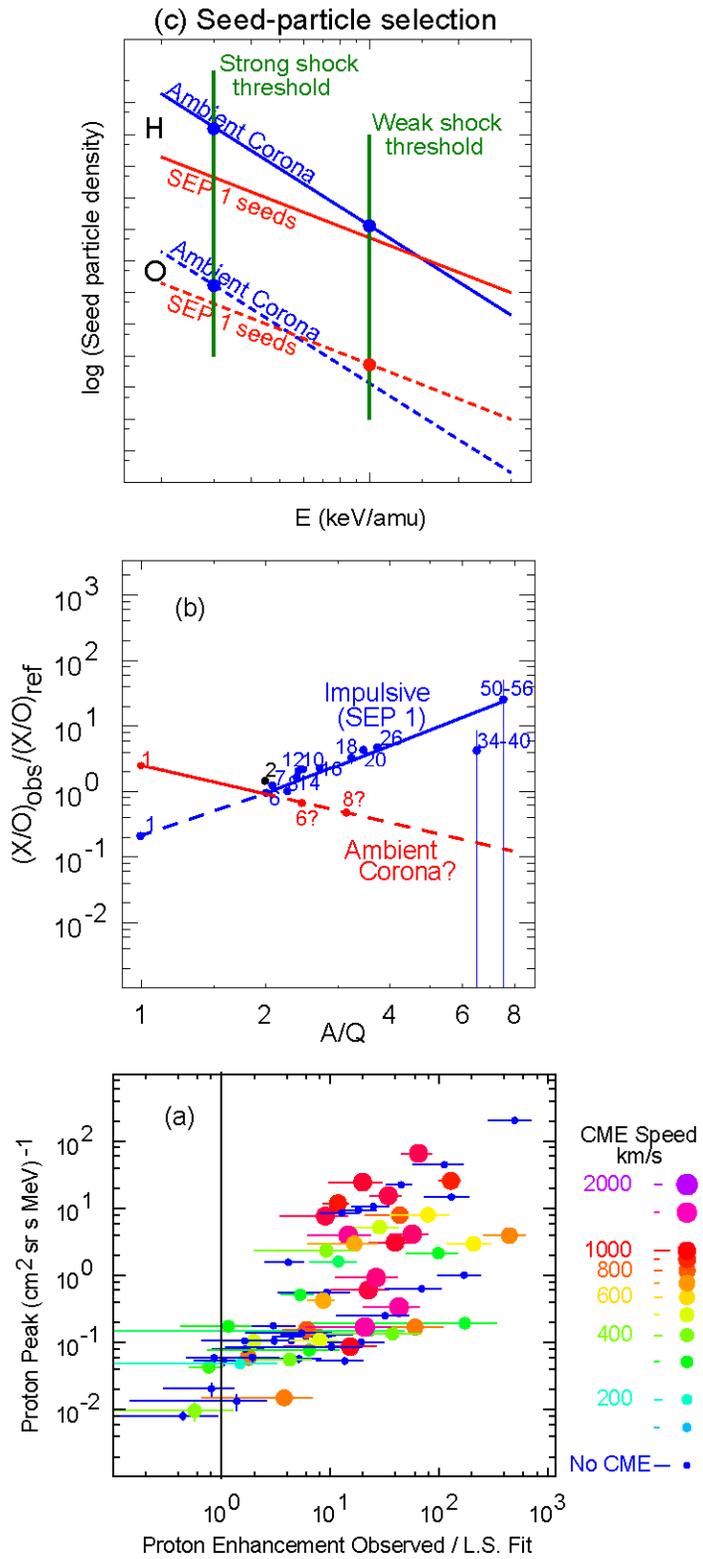

The selection of ions from the SEP1 seed population in Fig. 6(c) follows the same logical pattern as the selection of different Fe/O ratios suggested by Tylka et al. (2005). This dual-source selection provides more H than would be available from SEP1 ions alone, which have low H/O. We will see that this is not the only way to generate a proton excess but it seems the most likely way for weak or quasi-perpendicular shocks, including the relatively weak shocks in jets. Weaker shocks probably require higher injection energy (Tylka et al. 2005), although the "injection problem" (Zank et al. 2001) is not yet solved.

### *3.3 Smaller Impulsive Events*

Impulsive SEP events are small and often occur in a high proton background. Only 70 of the 111 impulsive SEP events listed by Reames et al. (2014a) have measurable proton intensities above background (Reames 2019b). Figure 6(a) suggests that the proton excess and proton peak intensities help distinguish SEP1 and SEP2 events. Of course, an associated CME speed > 500 km s$^{-1}$ is the best evidence of a SEP2 event in any case, but high intensities of He or O are also weak proxies if CME and proton data are absent.

Our criteria of using Fe/O to define impulsive events (see Fig 3(a)) have been justified above. However, it is possible to find many more smaller impulsive SEP events by using $^3$He/$^4$He. This is because there are about 50 times as many $^4$He ions as O ions in impulsive events, so we can get a statistically significant number of He ions without seeing any *Z* > 2 ions. Some of these smaller events have been included in previous studies (e.g. Nitta et al. 2006) but protons were not considered there, and, of course, power-laws in *A/Q* could not be studied for these small events. Nevertheless, it might be possible to study other impulsive event properties using these more-numerous small $^3$He-rich events.

### *3.4 Flares*

Flares branch from the path that leads to impulsive SEP events at Fig. 1(c). Magnetic reconnection in flares and jets would be expected to produce similar ion acceleration. Energetic ions on magnetic loops in flares eventually scatter into the loss cone producing nuclear reactions in the low corona that emit neutrons (Evenson et al. 1983, 1990) and γ-ray lines (e.g. Ramaty and Murphy 1987; Kozlovsky, Murphy, and Ramaty 2002) in the footpoints of the magnetic loops, eventually leading to hot, bright plasma filling the loops. Analysis of γ-ray lines from large flares has suggested that the accelerated "beam" is both $^3$He-rich (Mandzhavidze et al. 1999; Murphy, Kozlovsky, and Share 2016) and Fe-rich (Murphy et al. 1991), just like the impulsive SEP events that escape into space from jets.

However, flares exist, and they are bright and hot because all of the energetic particles accelerated in them are magnetically trapped and do *not* escape into space. The plasma reaches temperatures of 10 – 40 MK. There is no evidence of either hot plasma or nuclear reaction products in the SEPs we observe in space. Reconnecting closed field lines with other closed lines cannot directly produce open fields (Reames 2002).

7 February 2020                                                                                           D. V. Reames## 4 Gradual SEP Events

For gradual events, the value of Fe/O, averaged over the event, is less than four times the coronal value of 0.131, as seen in Fig. 3(a). SEP events selected in this way are large, persistent events associated with wide, fast CMEs (Kahler et al 1984). Owing to their long durations and high intensities we can analyze each of them in several 8-hr intervals; for magnetically well-connected events, differences in ion transport causes Fe/O to be enhanced early and decline to an Fe-depleted value later, with the *A/Q*-dependence also following this trend. We can only determine temperatures when the *A/Q*-dependence of the abundance enhancements is either rising or falling, not when it is flat. It has been gratifying that we usually find similar estimates of temperature in both the rising and falling phases of an event.

Perhaps it is surprising that as many as 25% of gradual events, listed by Reames (2016a), have steeply ascending *A/Q*-dependence that persists throughout the events with source plasma temperatures of ≈3 MK. When we extend the power-law fit lines to H we find that all of these events also have proton excesses just like the SEP2 events (Reames 2019c).

Shock waves continue to accelerate ions far out from the Sun. However, during this journey, they continue to reaccelerate the intense energetic ions that they originally selected from the denser seed populations very early in the event when they were much stronger and closer to the Sun, typically beginning at 2 – 3 solar radii (Reames 2009a, 2009b, Cliver et al. 2004). This onset occurs after the shock speed exceeds the local Alfvén speed, and the shock exits any significant magnetic loops, except possibly from preexisting structures like streamers (Kong et al. 2017). It is ironic that plasma destined to become SEPs must enter the corona on closed loops to produce the right FIP pattern, but later must be accelerated on open field lines to escape the Sun as SEPs we can measure.

### *4.1 Weak Shocks, T ≈ 3 MK, SEP3*

Figure 7 shows an analysis of the gradual SEP event of 18 April 2014. This event with *T* ≈ 2.5 MK, and rising *A/Q*-dependence up to an intensity peak at the local shock passage on 20 April, has the signature of typical SEP1 impulsive suprathermal ions with the persistent proton excess we have just associated with shock-reaccelerated SEP2 events. For the last time interval analyzed, on 20 April, the power law is too flat to define an accurate value of *T*. The high intensities and duration of several days already tends to distinguish this event from impulsive events. CME observations show an associated halo CME with a moderately high shock speed of 1203 km s$^{-1}$ (Reames 2016a). We will show a comparison of relative shock strengths later, but for now it seems these gradual SEP3 events derive their proton excess from a dual seed population, like the SEP2 events described in Fig. 3(b) and Fig. 3(c). The seed population for SEP3 events could also contain some SEP2 ions with a built-in proton excess.

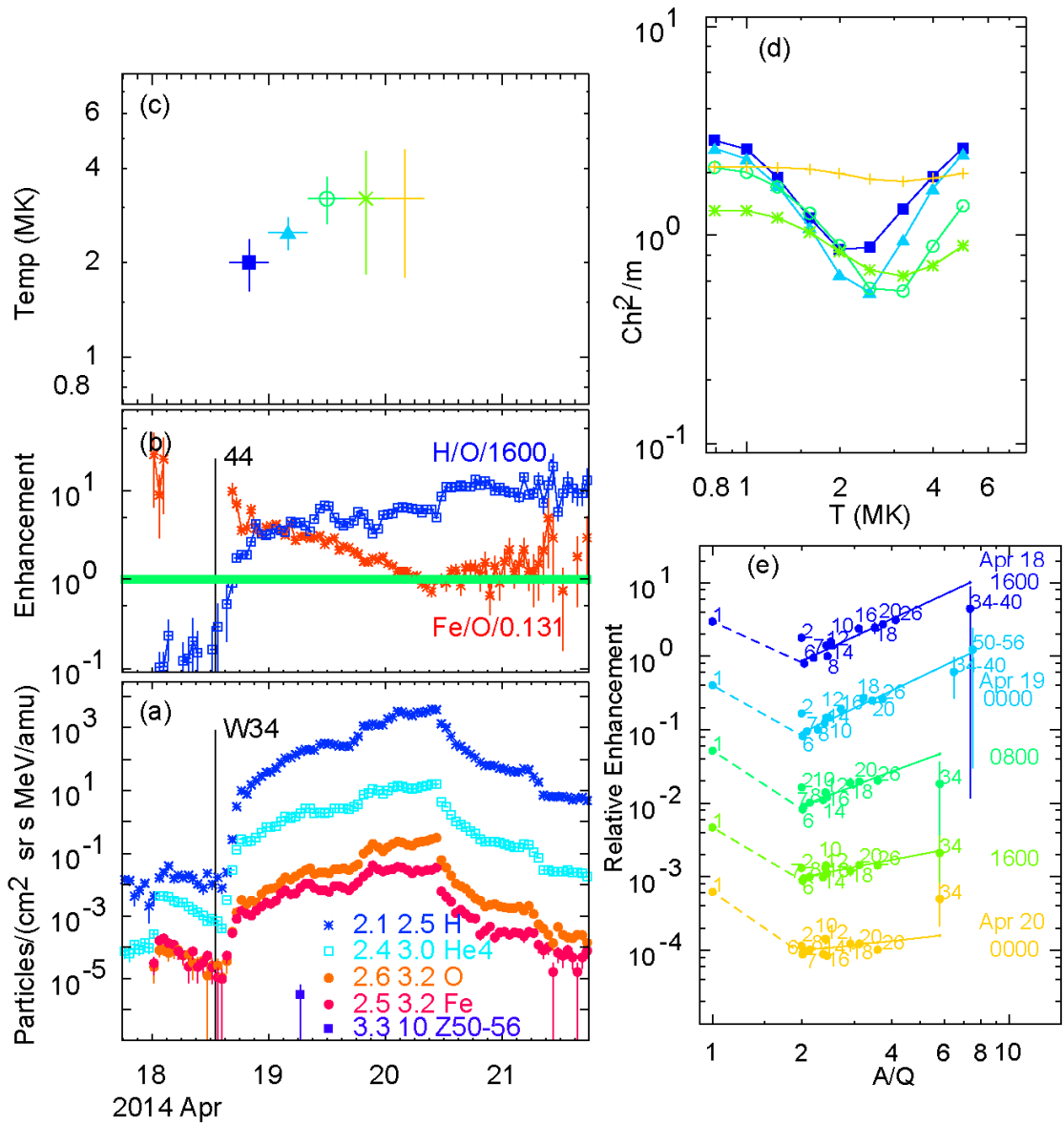

**Fig. 7** Panel (**a**) shows selected particle intensities and (**b**) shows abundance enhancement ratios for the SEP3 event of 18 April 2014. Panel (**c**) shows color-coded best-fit temperatures vs. time while panel (**d**) shows $\chi^2/m$ vs. $T$ for each time interval and panel (**e**) shows best-fit enhancement vs. $A/Q$ for each time interval displaced ×0.1, with numbers indicating $Z$ of the ions (Reames 2019c). The shock in this event samples SEP1 seed particles; even a few $34 \leq Z \leq 40$ and $50 \leq Z \leq 56$ ions persist.

The important difference between the SEP2 and SEP3 events is shown in Fig. 8. We envision a SEP2 event with a CME-driven shock from a single local jet reaccelerating the SEP1 ions from that same jet, as shown in Fig. 8(a). In Fig. 8(b), a wide, fast, CME-driven shock samples suprathermal residue from many (N) previous SEP1 and SEP2 events producing average abundances with variations reduced by a factor of $\sqrt{N}$ (Reames 2019d). He, C, and O are likely to be fully ionized at $T \approx 3$MK so these variations probably do not come from $Q$ variations, they must exist in the local source plasma.

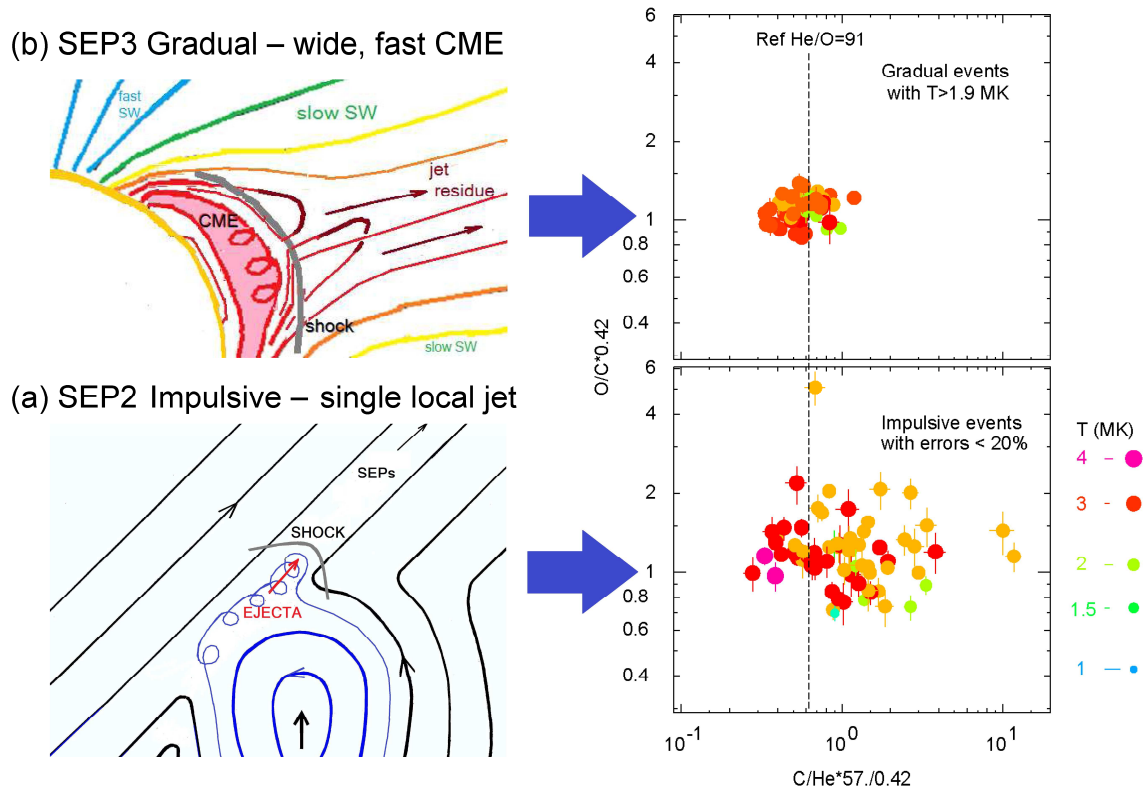

**Fig. 8** Panel (**a**) shows a local CME-driven shock wave from a single jet that reaccelerates SEP1 ions to produce an SEP2 event. Abundances vary from event to event as shown to the *right*. In (**b**) the shock from a wide, fast CME is poised to sweep up residue from many precious SEP1 or SEP2 jets, generating averaged abundances with much smaller variations in the corresponding distribution to the *right*. At $T \approx 3$ MK, He and C are fully ionized and should not participate in *A/Q*-dependent enhancements, only event-to-event source fluctuations (Reames 2019d).

For many years we have known that summing over "quiet" period between SEP events near solar maximum will produce $^3$He-rich, Fe-rich material (Richardson et al. 1990), even above 1 MeV amu$^{-1}$. There is a background of SEP1 ions from events that are too small to resolve individually. There are long periods of strong and recurrent enhancements in $^3$He/$^4$He and Fe/O that are now seen in otherwise quiet periods (Desai et al. 2003; Bučík et al. 2014, 2015; Chen et al. 2015), but only when an active region is magnetically connected during very quiet times (Mason et al. 2009).

As we proceed to smaller and smaller flares, their number increases as a power law (e.g. Lin et al. 1984), leading Parker (1988) to suggest that "nanoflares" are sufficiently numerous and energetic to heat the corona. Solar jets, as the open-field version of flares, are likely to have similar size behavior, and, while we do not know how the acceleration scales with size, it may be "microjets" or "nanojets" that provide the often-observed SEP1 seed particles for SEP3 events. Whatever we label these small jets, a substantial background population of SEP1 ions is observed to be available for shock acceleration and the process of averaging over a large number of individual sources reduces the variations of the abundances in SEP3 events. To reduce the 30% spread of abundances of the single-jet SEP1 and SEP2 events to the 10% spread of an SEP3 events requires averaging over the remnants of ≈10 single jets of comparable size. Since He and C are fully ionized at 3 MK, variations in C/He, for example, are not caused by *A/Q*.

## *4.2 Strong Shocks, T < 2 MK, SEP4*

An abundance analysis of two large western gradual events is shown in Fig. 9. These events come from the "Halloween" series of events, and the second event on 28 October 2003 is a GLE (Cliver 2006; Gopalswamy et al. 2012; Mewaldt et al. 2012).

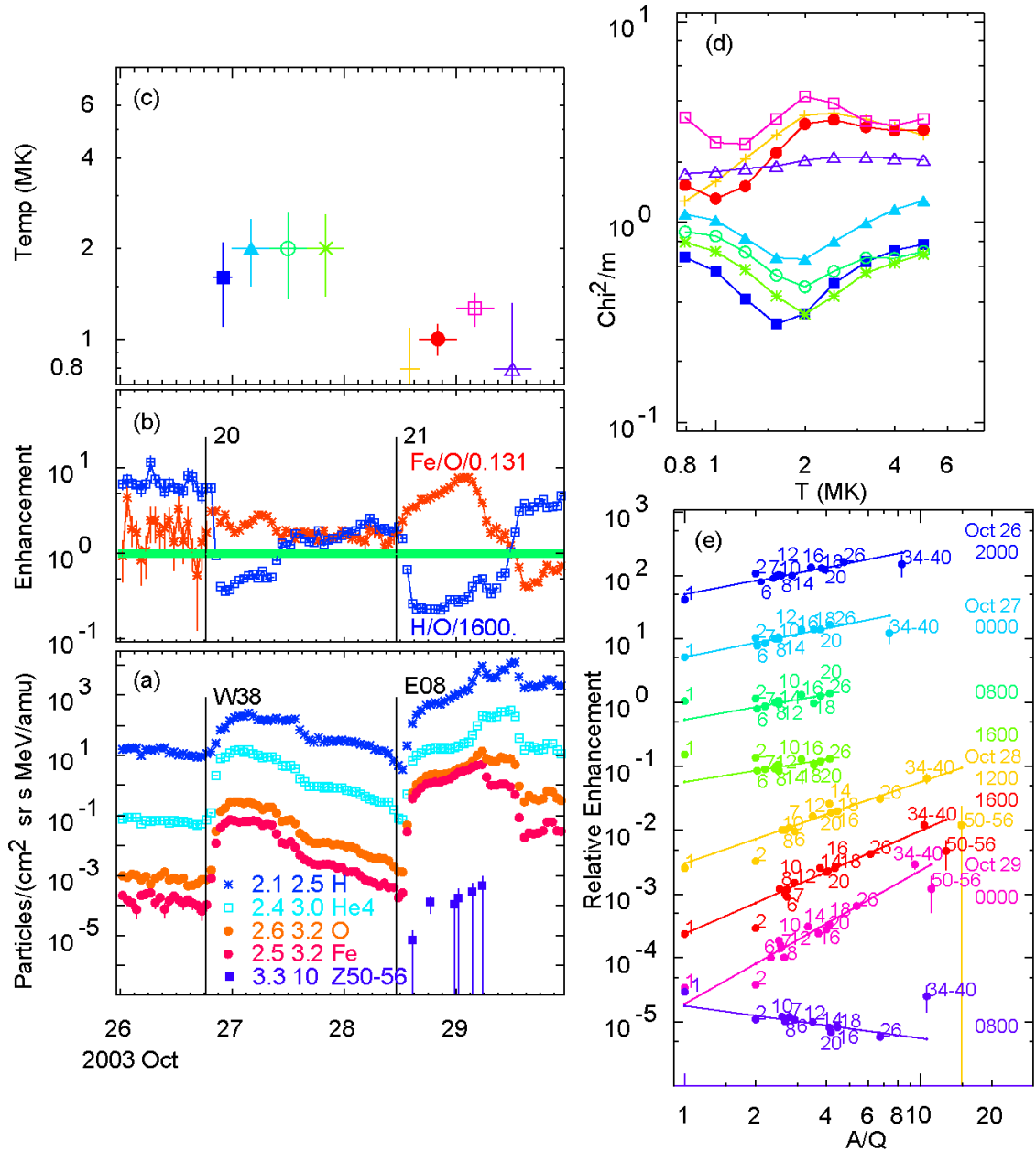

**Fig. 9** Panel (**a**) shows selected particle intensities and (**b**) shows abundance enhancement ratios for the two SEP4 events of 26 and 28 October 2003. Panel (**c**) shows color-coded best-fit temperatures vs. time while panel (**d**) shows $\chi^2/m$ vs. $T$ for each time interval and panel (**e**) shows best-fit enhancements vs. $A/Q$ for each time interval displaced ×0.1, with numbers indicating $Z$ of the ions (Reames 2019c). Proton enhancements at $A/Q = 1$ fall close to the fit lines from $Z > 2$ in all periods in (**e**).

The power-law fits of the ions with $Z > 2$ in Fig. 9(e) do well at predicting the intensity of H throughout the two events studied, even when the slope switches from ascending to descending with $A/Q$. These are both large events and the associated CME speeds are 1537 and 2459 km s$^{-1}$, respectively. However, many large gradual SEP events, especially eastern events, have declining power-law dependence on $A/Q$, like the event shown in Fig. 10. Again, the proton enhancements fall close to the $Z > 2$ fit lines in Fig. 10(e).

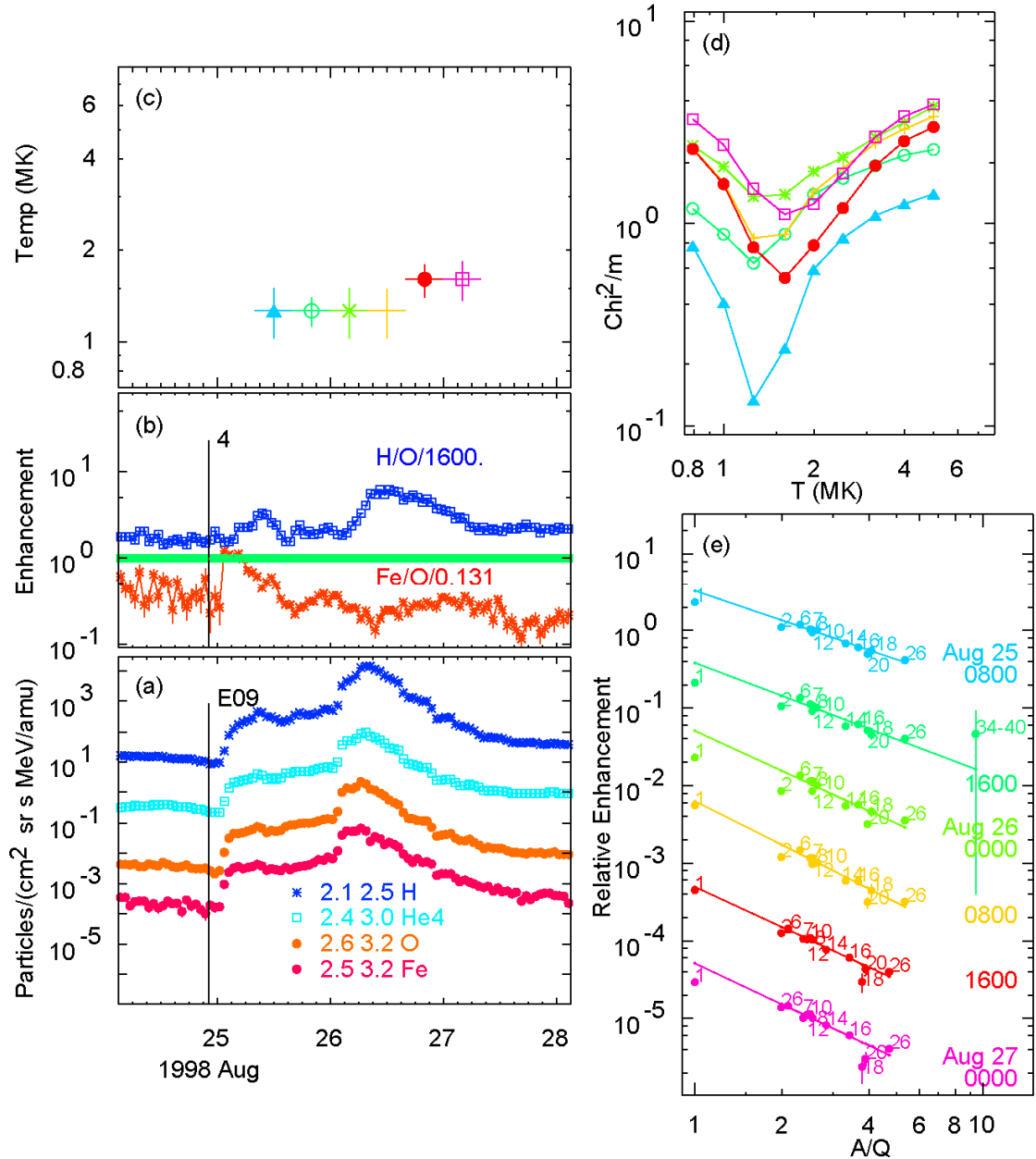

**Fig. 10** Panel **(a)** shows selected particle intensities and **(b)** shows abundance enhancement ratios for the event of 24 August 1998. Panel **(c)** shows color-coded best-fit temperatures vs. time while panel **(d)** shows $\chi^2/m$ vs. $T$ for each time interval, and panel **(e)** shows best-fit enhancements vs. $A/Q$ for each time interval displaced ×0.1, with numbers indicating $Z$ of the ions (Reames 2019c). Proton enhancements at $A/Q = 1$ fall close to the fit lines from $Z > 2$ in all periods in **(e)**.

Finally, however, a counter example in Fig. 11 shows an event which has proton excesses early (Fig. 11(e)) that vanish later in the event.

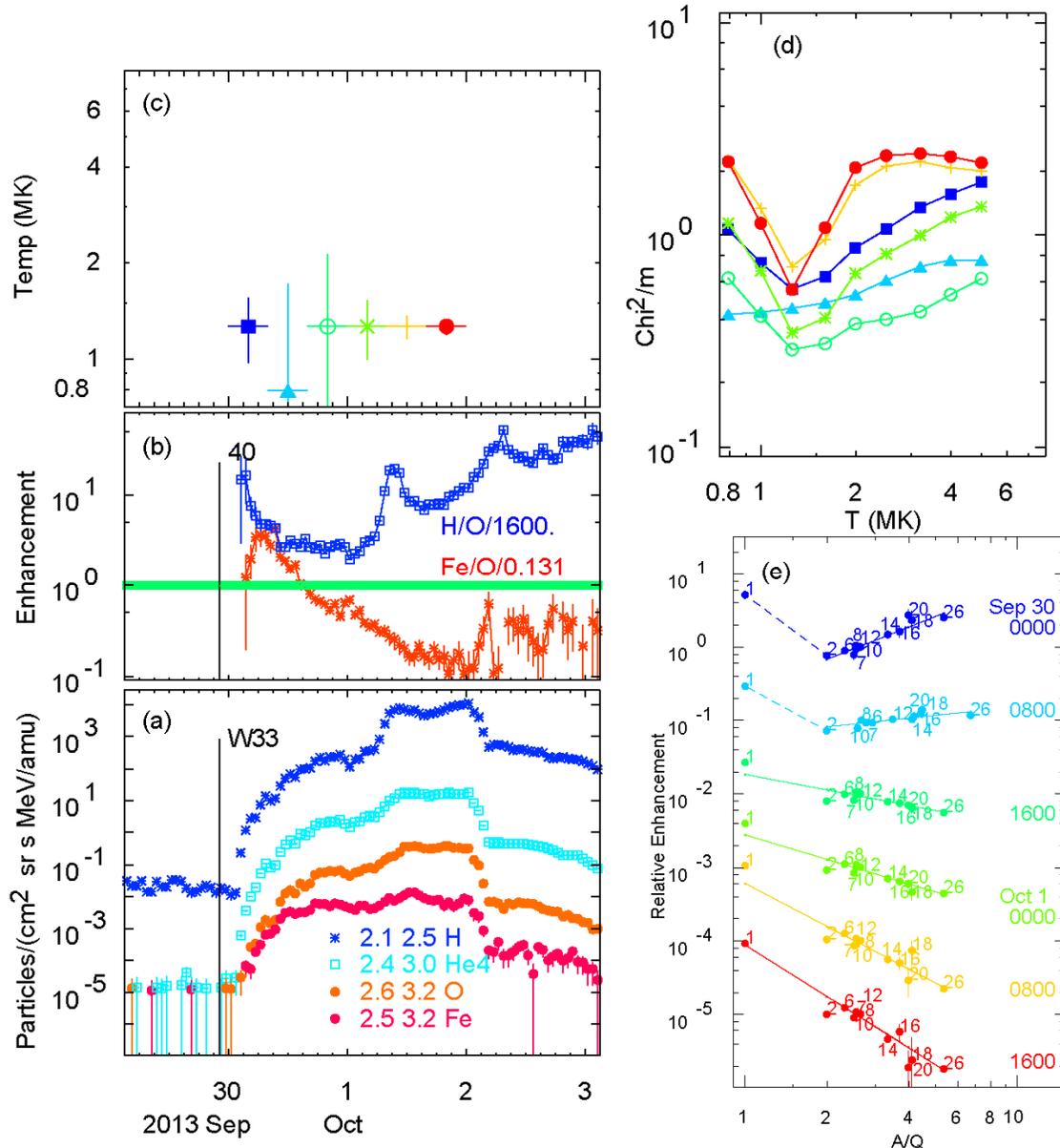

**Fig. 11** Panel **(a)** shows selected particle intensities and **(b)** shows abundance enhancement ratios for the SEP4 event of 29 September 2013. Panel **(c)** shows color-coded best-fit temperatures vs. time while panel **(d)** shows $\chi^2/m$ vs. $T$ for each time interval, and panel **(e)** shows best-fit enhancements vs. $A/Q$ for each time interval displaced ×0.1, with numbers indicating $Z$ of the ions (Reames 2019c). Early proton excesses cannot be explained by two-component seed population.

The event in Fig. 11 follows the classic pattern for a western event where Fe/O is enhanced early but depleted later because Fe scatters less than O, and the behavior of the power-law fit follows this pattern. However, proton excess cannot be explained by the two-component seed-population model because the temperature does not indicate the presence of any SEP1 ions; i.e. the $Z > 2$ ions appear to have the same temperature as the ambient plasma, early and late in the event.

The power-law fits we are using for gradual events are based upon the approximation that the scattering mean free path has a power-law dependence on magnetic rigidity. This is not necessarily the case early in large gradual SEP events when intensities of streaming protons are adequate to amplify Alfvén waves and alter their own transport, causing wave spectra to vary in time and space (Ng et al. 1999, 2003). The wave number of resonant waves is $k \approx B/\mu P$ where $P=pc/Qe$ is the rigidity of a particle of charge $Qe$, and momentum $p$, and $\mu$ is the cosine of its pitch angle relative to $B$. For example, 10-MeV protons propagate out early with $\mu \approx 1$ to generate resonant waves that scatter our 2.5-MeV amu$^{-1}$ He and other elements with $A/Q \geq 2$, delaying their arrival, but the 2.5 MeV protons arriving early with $\mu \approx 1$ encounter no increase in the ambient pre-event waves; this early increase and subsequent decrease in H/He (also in H/O) has been studied previously (Reames et al. 2000) and is seen in Fig. 11(b) but not the events in Figs. 9(b) or 10(b), which have high pre-event proton background. Since the waves are amplified almost entirely by protons the break in the behavior comes between H and He. Intense SEP4 events break the power-law rules, especially early in the events.

Figure 12 shows the distributions in CME speed of the SEP3 and SEP4 events. The mean CME speed for the SEP3 events is 1250 km s$^{-1}$ and that for SEP4 events is 1764 km s$^{-1}$. The SEP3 events tend to be found at the beginning of the solar cycle and they dominate the weak Solar Cycle 24.

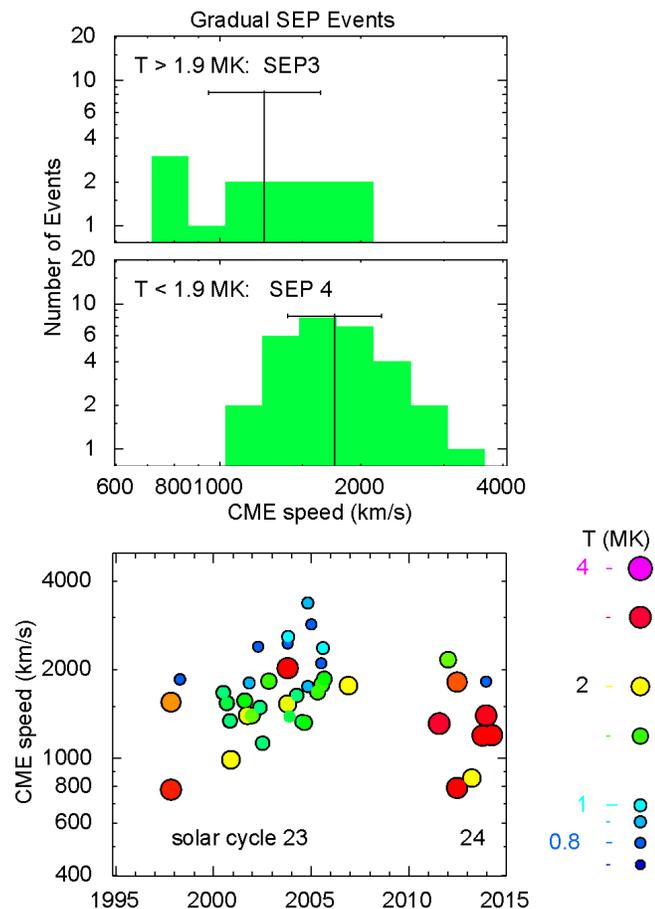

**Fig. 12** The distributions of CME speed is shown for SEP3 events (*upper panel*) and SEP4 events (*middle panel*). The *lower panel* shows the dates of SEP events with different associated CME speed with the source plasma temperature indicated by the point size and color. Unfortunately, these CME speeds are not usually determined for the same magnetic flux tube as the SEP abundances. (Reames 2019d)

While the proton excess or suppression may be dependent on the time evolution of the wave spectrum, and thus may occasionally increase or decrease in large gradual events, these SEP4 events are clearly distinguished from the SEP3 events by their acceleration of ambient plasma with $T < 2$ MK.

### *4.2 Undefined Temperatures*

Temperature measurements do play a role in distinguishing SEP3 and SEP4 events. The list we have studied contains 45 gradual events for which temperatures have been assigned (Reames 2016a) which came from a list of 62 candidate events. This event selection required at least 4 reasonably consistent 8-hr temperature measurements. The most common problem with the 17 rejected events was that the abundance enhancements were too flat to define a temperature, i.e. the observed abundances were so similar to the reference coronal abundances in Table 1 that no preferred dependence upon *A/Q* could be found. There is nothing wrong with these events. These events with undefined temperatures are undoubtedly mostly SEP4 events because they clearly lack the strong ascending power-law dependence at $T \approx 3$ MK required of SEP3 events. Temperature measurements are not strictly necessary to identify the source of the dominant shock-accelerated seed population.

## 5 Comparisons

Having defined all four event classes it seemed appropriate to try to exhibit and compare all four in a single figure. The left side of Fig. 13 shows impulsive SEP events and the right side gradual SEP events. The lower panels, (a) and (d), show representative particle intensities at the same scales of time and intensity, central panels show the best-fit source-plasma temperatures for the times shown, and the upper panels show the best power-law fits vs. *A/Q* for each time interval, color coded with the temperature below.

We have listed impulsive Event 96 as a possible SEP1 event, even though it has some proton excess, because the excess is barely one standard deviation away from the fit line. Of course we cannot really exclude some reacceleration by a shock, but no CME is associated with Events 96 or 97. Perhaps we have failed to include a pure SEP1 this time, but this event sequence is quite interesting in its own right. Event 98 has a 1044 km s$^{-1}$ CME associated and a large proton excess; it is clearly a SEP2 event. Note that the proton excess seems to increase systematically from Event 96 to 97 to 98.

Another unusual feature of the impulsive events in Fig 13(c) is the strong He suppression in Event 96. This He suppression seems to diminish in Events 97 and almost disappears in Event 98 when the full measure of He finally arrives. The apparent coupling among these three events is remarkable and of unknown origin; the proton excess increases while the He suppression decreases. We should caution, however, that Events 34 and 35 (Reames 2019b) have the opposite behavior, the proton excess decreases and the He suppression increases with time, so any apparent coupling may be coincidental. Note also that Event 14, shown in Fig. 4(e), shows some suppression of He while the other four events in that figure show little.

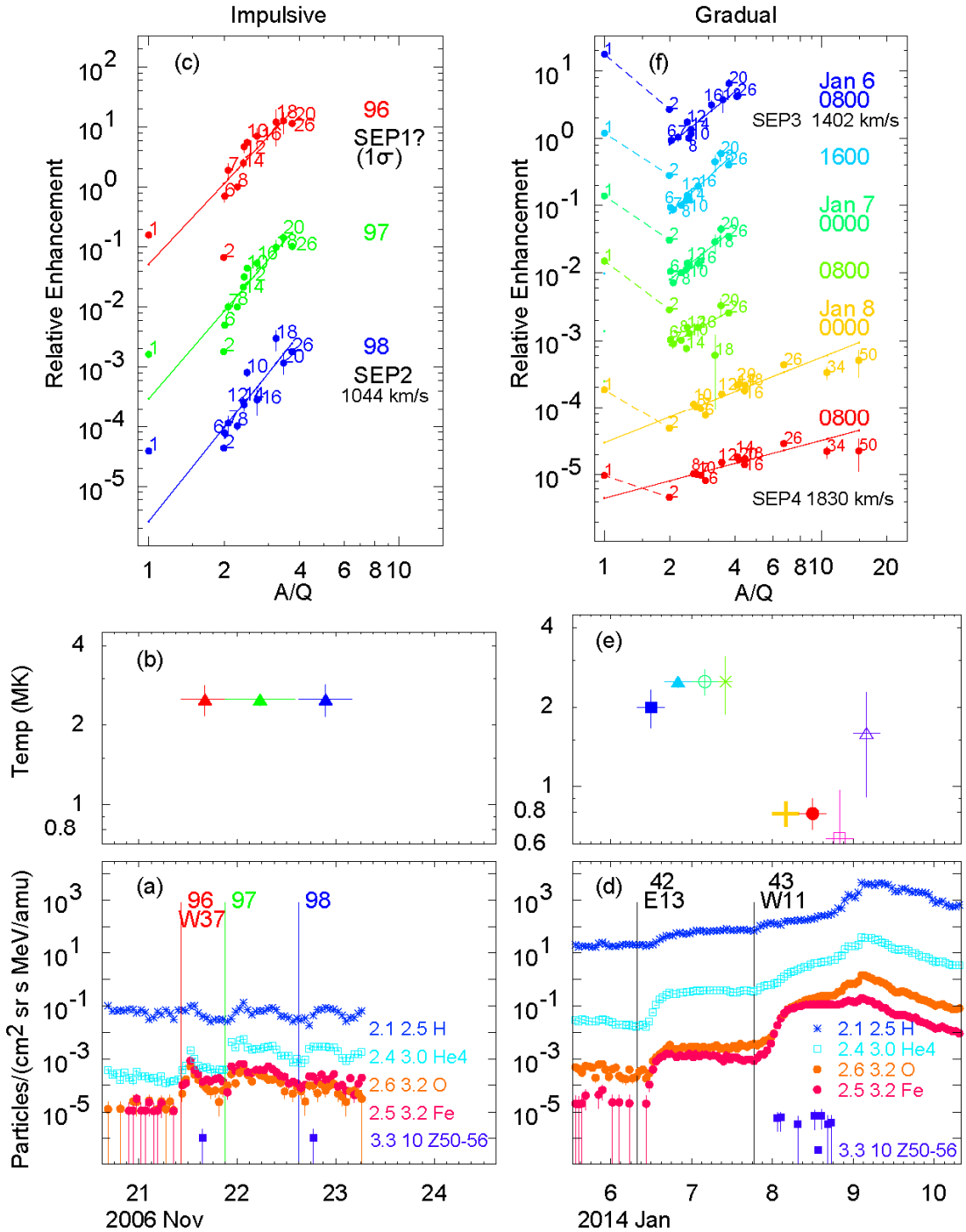

**Fig. 13** The *lower* panels compare same particle intensities for (**a**) impulsive and (**d**) gradual SEP events at the same scale. Onsets times are flagged with event numbers for impulsive (96, 97, 98, from Reames et al. 2014a) and gradual events (42, 43, from Reames 2016a). *Central* panels, (**b**) and (**e**), show fitted source plasma temperatures, vs. time, for measured intervals, and *upper* panels, (**c**) and (**f**) show enhancements vs. *A/Q* for each interval, color coded. See text.

The two gradual events on the right in Fig. 13 are well defined. Event 42 is a SEP3 event with a source temperature of 2.5 MK throughout and a strong proton excess, despite the pre-event proton background. Event 42 even appears to have a rare He excess, which is probably also accelerated from the ambient plasma, like the protons.

Event 43 is a clear SEP4 with $T \approx 0.8$ MK. The small proton excess may all be background from the previous event. The associated CME speed for Event 42 is 1402 km s$^{-1}$ and that for Event 43 is 1830 km s$^{-1}$, i.e. the SEP3 event has a weaker shock than the SEP4.

## 6 Discussion

Measuring power-law abundance patterns is a useful way to estimate source plasma temperatures in SEP events and to distinguish the signature of ≈3 MK plasma of impulsive-SEP, i.e. SEP1 material, an imprint originally produced by magnetic reconnection in solar jets. Proton excesses combined with these SEP1 abundances generally reveal the additional action of shock acceleration on two-component seed populations to produce SEP2 or SEP3 components. However, proton excesses can also be produced by changes in the scattering produced by proton-generated waves that disrupt the simple power-law behavior in some large gradual or SEP4 events. Wave generation in SEP3 events would mostly enhance the two-component proton excess that is already present.

SEP3 and SEP4 components are easily distinguished by their dominant source plasma temperatures. In comparing CME speeds of these two populations there is always the unfortunate issue that the CME speeds are not determined along the same magnetic flux tube as the SEP abundances are measured. Also, we have not been able to determine $\theta_{Bn}$ near the region of early acceleration near the Sun. The work of Koulomvakos et al. (2019) suggests a possible remedy for this problem.

It is quite likely that all gradual SEP events have a component of reaccelerated ions from a SEP1 seed-population source. In SEP3 events they actually dominate, but the SEP4 events with a small component of $^3$He found by Mason et al. (1999) suggest the commonality of small SEP1 contributions. Measurements of charge state distributions could tell us how much impulsive plasma was present, especially if the SEP1 ions were highly stripped as observed by DiFabio et al. (2008), but unfortunately such measurements are uncommon, especially for ions >1 MeV amu$^{-1}$. It would be very interesting to know the percentage of SEP1 ions in gradual events and its dependence upon shock parameters near the base of the observer's magnetic flux tube.

If SEP1 ions are stripped after acceleration, as seems to be the case (Luhn et al. 1987; DiFabio et al. 2008), only the power-law abundances give us information on temperatures before or during acceleration, charge measurements near Earth do not. However, the change in the patterns of $Q$ and $A/Q$ caused by stripping mean that transport of the ions after shock reacceleration in SEP2 and SEP3 populations will actually involve different power-law patterns and effective temperatures after acceleration. Yet, we seem to find the same $T \approx 3$ MK for these populations as well. This is probably because the SEP1 power of $A/Q$ is so steep, compared with the modest increase or decrease during transport, that transport has little effect. For example, if acceleration in a jet at $T \approx 3$ MK has caused enhancements in the order Ne > Mg > Si, subsequent transport will be very unlikely to undo this ordering, especially if all three of these elements at 2 – 5 MeV amu$^{-1}$ have been fully ionized after acceleration so that they now have $A/Q = 2$ and their relative abundances can no longer change.  Similarly, the large enhancements of heavy ions will be only modestly enhanced or reduced further by transport after velocity-dependent shock acceleration.  The $\chi^2/m$ selection of temperature for the SEP3 event in Fig. 7(d) is just as clear as that for the SEP1 or SEP2 events in Fig. 4(d).

We have not discussed reacceleration of remnant suprathermal ions left over from gradual SEP events.  While this is possible during sequences of gradual SEP events, there may be a large number of nanojets available to produce SEP1 ions, but no corresponding population of common small gradual events.  Reaccelerated ions from gradual events would also be hard to distinguish from ambient plasma.

Suppression of He (Reames 2019a) seems to be extreme evidence of the incomplete ionization of He caused by its uniquely high FIP of 24.6 eV, which makes it the last element to be fully ionized during FIP processing (Laming et al. 2009).  FIP processing is usually considered to be slow and produces He abundances varying by a factor of 2 as observed in gradual events (Reames 2017b).  Does magnetic field motion in some jets suddenly separate incompletely-processed He ions from the residual neutral atoms?  If the He suppressions are a result of incomplete FIP processing, rather than acceleration, then they might also be found in the associated CMEs as well as in the SEPs.  Are there samples of solar wind associated with jets that have He/O < 10?  He suppression is actually fairly rare in SEP1 events, but it can be quite striking when it occurs.  Events with strong He suppression are some of the smallest events we have studied, with some of the steepest powers of $A/Q$.  These studies reveal interesting and extreme phenomena that are not clearly understood.

## 7 Summary

Plasma in the solar corona that may later become SEPs has a FIP-dependent pattern of element abundances, relative to that of the photosphere, which differs from plasma that will later become solar wind.  Theory tells us that the former enters the corona on closed magnetic field lines and the latter on open field lines.

Element abundances, relative to the coronal abundances, define a paradigm with four distinct populations of SEPs:

- SEP1 ions are accelerated in islands of magnetic reconnection in solar jets. Relative abundance enhancements increase 1000 fold as a steep power law in $A/Q$ at $T \approx 3$ MK from H to Pb with powers ranging from 3 to 6.  There may also be 1000-fold enhancements of $^3$He/$^4$He. Any CMEs from these jets do not drive shock waves fast enough to re-accelerate the SEPs.

- SEP2 ions arise in jets with CMEs that drive shock waves fast enough to re-accelerate the SEP1 ions along with excess protons from the ambient plasma. The steep power law of the SEP1 ions and $T \approx 3$ MK imprint persist for the ions with Z > 2.

- SEP3 ions are accelerated along the front of shock waves driven by wide, fast CMEs which preferentially sweep up residual SEP1 suprathermal ions, left by a large number of jets around an active region, along with an excess of protons from the ambient plasma. Quasi-perpendicular shock waves may also show this dominance of pre-accelerated SEP1 ions. The steep power law of the SEP1 ions and $T \approx 3$ MK signature largely persist for $Z > 2$ ions, but abundance variations are reduced by averaging over the multi-jets source.

- SEP4 ions are accelerated from the ambient plasma along extremely strong shock waves driven by wide, fast CMEs. Protons usually fit the same power law in $A/Q$ as $Z > 2$ ions with no proton excess and the events have $0.8 < T < 2.0$. In some events wave-generation by streaming protons can disrupt the simple power law, especially early in the event. Any SEP1 ions in SEP4 events are buried.

These categories of events are identified by the behavior of the simple assumed power-law dependence of abundances on $A/Q$. The relative abundance of H and the persistent characteristic behavior of the impulsive SEP1 pattern play special roles in our understanding. Events still fall into the categories of impulsive and gradual events, but each of those categories has split. SEP1 and SEP2 are distinguished in Fig. 6. SEP2 and SEP3 are distinguished in Fig. 8. SEP 3 and SEP4 are distinguished in Fig 12 and Fig 6(c). The principle SEP abundance properties are summarized in Table 2.

**Table 2** Properties of the Four SEP Abundance Patterns.

|      | **Observed Properties** | **Physical Association** |
|------|------------------------|--------------------------|
| **SEP1** | Power law enhancement vs. $A/Q$ with $T \approx 3$ MK for $Z=1$ and $Z >2$ | Magnetic reconnection in solar jets with no fast shock |
| **SEP2** | Power law enhancement vs. $A/Q$ with $T \approx 3$ MK for $Z >2$ ~30% scatter in He/C, etc. Proton excess | Jets with fast, narrow CMEs drive shocks that reaccelerate SEP1 ions plus excess protons from ambient plasma |
| **SEP3** | Power law enhancement vs. $A/Q$ with $T \approx 3$ MK for $Z >2$ <10% scatter in He/C, etc. Proton excess | Moderately fast, wide CME-driven shocks accelerates SEP1 residue left by many jets, plus excess protons from ambient plasma |
| **SEP4** | Power law enhancement vs. $A/Q$ with $0.8< T < 1.8$ MK for $Z=1$ and $Z > 2$ | Extremely fast, wide CME-driven shocks accelerate all seed ions so that ambient plasma dominates. |